\documentclass{jfm}
\usepackage{graphicx}
\usepackage{epstopdf, epsfig}
\usepackage[]{psfrag}
\usepackage{xcolor}
\usepackage{url}

\shorttitle{Coherence of superstructures in turbulent Rayleigh-B\'{e}nard flow}
\shortauthor{D. Krug \emph{et al.}}

\title{Coherence of temperature and velocity superstructures\\ in turbulent Rayleigh-B\'{e}nard flow}

\author{Dominik Krug\aff{1}
 \corresp{\email{d.j.krug@utwente.nl}},
 Detlef Lohse\aff{1,2},
 \and Richard J.A.M. Stevens\aff{1}}

\affiliation{
\aff{1}Physics of Fluids Group and Twente Max Planck Center, 
Department of Science and Technology, Mesa+ Institute,
 and J.M. Burgers Center for Fluid Dynamics, University of Twente, P.O Box 217, 7500 AE Enschede, The Netherlands \\
\aff{2}Max Planck Institute for Dynamics and Self-Organization, Am Fassberg 17, 37077 G\"ottingen, Germany}

\newcommand{\blue}[1]{\textcolor{black}{#1}}

\begin{document}

\maketitle

\begin{abstract}
We investigate the interplay between large-scale patterns, so-called superstructures, in the fluctuation fields of temperature $\theta$ and vertical velocity $w$ in turbulent Rayleigh-B\'{e}nard convection at large aspect ratios. Earlier studies suggested that velocity superstructures were smaller than their thermal counterparts in the center of the domain. However, a scale-by-scale analysis of the correlation between the two fields employing the linear coherence spectrum reveals that superstructures of the same size exist in both fields, which are almost perfectly correlated.  The issue is further clarified by the observation that in contrast to the temperature, and unlike assumed previously, superstructures in the vertical velocity field do not result in a peak in the power spectrum of $w$. The origin of this difference is traced back to the production terms of the $\theta$- and $w$-variance. 
These results are confirmed for a range of Rayleigh numbers $Ra = 10^5$--$10^9$, the  superstructure size is seen to increase monotonically with $Ra$.
Furthermore, the scale distribution of particularly the temperature fluctuations is pronouncedly bimodal. In addition to the large-scale peak caused by the superstructures, there exists a strong small-scale peak. This `inner peak' is most intense at a distance of $\delta_\theta$ from the wall and associated with structures of size $\approx 10 \delta_\theta$, where $\delta_\theta$ is the thermal boundary layer thickness.
Finally, based on the vertical coherence relative to a reference height of $\delta_\theta$, a self-similar structure is identified in the velocity field (vertical and horizontal components) but not in the temperature. 
\end{abstract}


\section{Introduction}
A remarkable feature of turbulent flows is that amid the inherent disorder both in time and space, they frequently give rise to a surprisingly organized flow motion on very large scales. Such very large-scale structures in the fully turbulent regime have, for example, been reported for turbulent boundary layers \citep{Hutchins2007}, plane Couette flow \citep{Lee2018}, and Taylor-Couette \citep{Huisman2014} turbulence. Here we focus on superstructures in turbulent Rayleigh--B\'enard convection (RBC), which is an idealized configuration that is widely used to study thermal convection \citep{Ahlers2009, Lohse2010, Chilla2012,Verma2018}. The strength of the non-dimensional thermal driving in RBC is given by the Rayleigh number $Ra$, while the dimensionless heat transfer is characterized by the Nusselt number $Nu$.

Large-scale organization in convective flows is widespread. An astonishing example is the formation of so-called cloud streets in the atmosphere that can extend for hundreds of kilometers \citep[e.g.][]{Young2002}. Studying related features in RBC requires a cell with a large aspect ratio $\Gamma$. Naturally, this poses a challenge to experiments and simulations. Experimentally \citep[e.g.][]{Fitzjarrald1976, Sun2005,sun05e, Zhou2012,hog13,duPuits2013, Cierpka2019}, it is very challenging to extract flow information beyond global parameters or local measurements of turbulence statistics. On the other hand, simulations for large aspect ratios are very costly if the thermal driving is sufficiently strong to achieve a moderately or even a strongly turbulent state. The first to tackle the problem numerically were  \citet{Hartlep2003}  and several related studies have since been presented in the literature \citep{Parodi2004,har05,shi05,shi06,shi07,Hardenberg2008,bai10,Emran2015,sak16}. Very recently,  the available ranges of large-aspect ratio simulations  have been  extended significantly in  Prandtl number ($Pr$) by \citet{Pandey2018} and in $\Gamma$ as well as in $Ra$ by \citet{stevens2018}.

\begin{figure}
\begin{center}
 \psfrag{c}{$\tilde{\theta}$}
{\includegraphics[width = \textwidth]{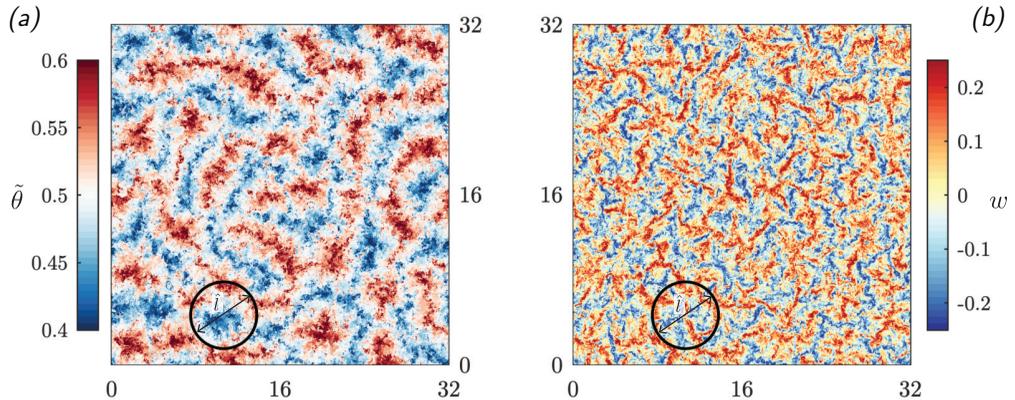}}
\caption{Snapshots of the temperature (a) and the vertical velocity (b) field at mid-height for a simulation in a $\Gamma = 32$ cell  with $Pr=1$ at $Ra= 10^8$. \textcolor{black}{The diameter of the circles in both panels indicates the  superstructure size $\hat{l} = 6.3$ (see table \ref{table1}).}}
\label{fig:snaps}
\end{center}
\end{figure}

From these papers, it has become clear that in RBC superstructures, i.e., \ flow structures that are significantly larger than the convection rolls at onset \citep[see, e.g.][]{Drazin2004} or in the weakly non-linear regime \citep{Morris1993}, exist at higher $Ra$. It is widely observed that the superstructure size increases with $Ra$ \citep{Fitzjarrald1976, Hartlep2003, Pandey2018, Green2019},  while the $Pr$-dependence appears to be more complicated. For the latter, \citet{Pandey2018} report that at $Ra = 10^5$ the largest structures are found for $Pr\approx 7$, but $Pr$-variations over a significant range at higher $Ra$ have not been reported yet. \citet{stevens2018} showed that very large domain sizes up to $\Gamma = 64$ are necessary to fully converge the size of the superstructures at $Ra = 10^8$. Finally, \citet{Hardenberg2008} and \citet{Pandey2018} demonstrate that superstructures evolve on timescales much longer than the free-fall time scale.

There is no consensus yet on how to best extract and quantify the superstructures in RBC. Researchers have relied on peaks in velocity and/or temperature power spectra \citep{Hardenberg2008,Pandey2018,stevens2018}, velocity-temperature co-spectra \citep{Hartlep2003,Green2019,Fitzjarrald1976} or so-called integral length scales \citep{Parodi2004,stevens2018} to determine the structure size. A puzzling and yet unexplained observation is that superstructures in the temperature ($\theta$) field are larger than in the vertical velocity $w$ \citep{Pandey2018,stevens2018} field when the structure size is determined based on the peaks in the power spectrum or the corresponding integral length scale. Also visually, the difference between the temperature and vertical velocity fields can easily be observed in the snapshots of the flow at mid-height, which are presented in figure \ref{fig:snaps}. This figure reveals that the vertical velocity field is dominated by significantly smaller structures than the temperature field. Moreover, the correlation between the two quantities appears much lower than one would naively expect, given that the temperature fluctuations provide the driving of $w$. These observations seem at odds with the notion that superstructures in RBC form large-scale convection rolls for which temperature and velocity scales should be of the same size.

To address and clarify this issue along with related questions, we use the dataset of \citet{stevens2018} to assess energy distributions and  coherence on a scale-by-scale basis. Before presenting our results in \S\ref{sec:res}, we provide the relevant details on the dataset of \citet{stevens2018}, together with the parameters of additional simulations performed for this study, in \S\ref{sec:dat}. We summarize our findings in \S\ref{sec:conc}.

\begin{table} 
\begin{center}
\begin{tabular}{cccccccccc}
\hline
$Ra$   & $N_{x}\times N_{y} \times N_{z}$  & $Nu$  &  $Re_h$   & $Re_v$   & $Re_t$  & $\hat{l}$&\blue{$\delta_\theta=1/(2Nu)$}  \\ \hline
$1\times10^{5}$ & $2048 \times 2048 \times 64$  & 4.35     & 55.7   &  40.3  &  68.7 &4.4 &\blue{0.115} \\ 
$4\times10^{5}$ & $2048 \times 2048 \times 64$  & 6.48     & 111.7   &   84.3 & 140.0 & 4.5 &\blue{0.077}\\ 
$1\times10^{6}$ & $3072 \times 3072 \times 96$  & 8.34     & 176.0   &  131.6  & 219.8 & 4.9& \blue{0.060} \\ 
$4\times10^{6}$ & $3072 \times 3072 \times 96$  & 12.27    &   349.7  &  250.8  & 430.4&5.4 & \blue{0.041} \\ 
$1\times10^{7}$ & $4096 \times 4096 \times 128$  & 15.85   &  547.1  &  380.1  &  666.2 & 5.9&\blue{0.032} \\ 
$1\times10^{8}$ & $6144 \times 6144 \times 192$  & 30.94 & 1660.3 & 1056.1 & 1967.8 & 6.3&\blue{0.016}  \\
$1\times10^{9}$ & $12288 \times 12288 \times 384$  & 61.83  & 4879.2 & 2962.3 & 5708.1 & 6.6&\blue{0.008} 
\end{tabular}
\caption{The columns from left to right indicate the $Ra$ number, the numerical resolution in the horizontal and wall normal directions ($N_{x}\times N_{y} \times N_{z}$), the $Nu$ number, and the horizontal \blue{($Re_h = \sqrt{\langle v_x^2+v_y^2 \rangle_V }\sqrt{Pr/Ra}$)}, vertical \blue{($Re_w=\sqrt{\langle w^2\rangle_V }\sqrt{Pr/Ra}$)}, and total \blue{($Re_t=\sqrt{\langle v_x^2+v_y^2+w^2\rangle_V }\sqrt{Pr/Ra}$)} Reynolds numbers. The length scale $\hat{l}$ denotes the superstructure scale based on the coherence spectrum $\gamma^2_{\theta w}$ (plotted as triangles in figure \ref{fig:Racoh}b) \blue{ and $\delta_\theta $ is the thermal boundary layer thickness. } }
\label{table1}
\end{center}
\end{table}

\section{Dataset} \label{sec:dat}
We solve the Boussinesq equations with the second-order staggered finite difference code AFiD. The code has been extensively validated and details of the numerical methods can be found in \cite{ver96,ste10,ste11,poe15c,zhu18b}. The governing equations in dimensionless form read:
\begin{eqnarray}
\frac{\partial \bf{u}}{\partial t} +{\bf u}\cdot \nabla {\bf u}&=&-\nabla p +\sqrt{\frac{Pr}{Ra}}\nabla^2 {\bf u}+\theta \hat z,\\
\nabla \cdot {\bf u}&=&0,\\
\frac{\partial \theta}{\partial t} +{\bf u}\cdot \nabla \theta&=&\frac{1}{\sqrt{RaPr}}\nabla^2 \theta,
\end{eqnarray}
where $\hat z$ is the unit vector pointing in the opposite direction of gravity, $\bf u$ the velocity vector normalized by the free fall velocity $\sqrt{g \alpha \Delta H}$, $t$ the dimensionless time normalized by $\sqrt{H/(g\alpha\Delta)}$, $\theta$ the temperature normalized by $\Delta$, and $p$ the pressure normalized by $g\alpha\Delta/H$. The control parameters of the system are $Ra=\alpha g \Delta H^3/(\nu \kappa)$ and $Pr=\nu/\kappa$, where $\alpha$ is the thermal expansion coefficient, $g$ the gravitational acceleration, $\Delta$ the temperature drop across the container, $H$ the height of the fluid domain, $\nu$ the kinematic viscosity, and $\kappa$ the thermal diffusivity of the fluid. The boundary conditions on the top and bottom plates are no-slip for the velocity and constant for the temperature. Periodic conditions in the horizontal directions are used. In all our simulations, $Pr$ is fixed to 1 and we analyze data for $\Gamma = L/H=32$, where $H$ is the vertical distance between the plates and $L$ the horizontal extension of the domain. \blue{Length scales are normalized by $H$ unless specified otherwise and we set $H = 1$.} Coordinates in the wall-parallel direction are denoted by $x$ and $y$ while the $z$-axis points along the wall-normal. Horizontal velocity components are denoted  $v_x$ and $v_y$, respectively.
A high spatial resolution in the boundary layer and bulk has been used to ensure that the resolution criteria set by \cite{ste10} and \cite{shi10} are fulfilled. Details about the simulations can be found in table \ref{table1}. The simulations for  $Ra=10^8$ and  $Ra=10^9$ have been reported before in \citet{stevens2018}, while the simulations for $10^5\leq Ra \leq 10^7$ have been performed for this study. The horizontal, vertical, and total Reynolds numbers indicated in table \ref{table1} represent the volume and time averages  of $Re_h = \sqrt{\langle v_x^2+v_y^2 \rangle_V }\sqrt{Pr/Ra}$, $Re_w = \sqrt{\langle w^2 \rangle_V }\sqrt{Pr/Ra}$, and  $Re_t=\sqrt{\langle v_x^2+v_y^2+w^2\rangle_V }\sqrt{Pr/Ra}$, respectively.  In the following, we decompose instantaneous quantities $\tilde{\psi} $ into mean and fluctuating parts according to $\tilde{\psi} = \Psi +\psi$, where $\Psi = \langle \tilde{\psi} \rangle$ with $\langle \cdot \rangle$ denoting an average over a wall-parallel plane and time.

\section{Results} \label{sec:res}

In presenting our results, we initially (\S\ref{sec:coh}--\S\ref{sec:cohspat}) restrict the discussion to a single Rayleigh number ($Ra = 10^8$). A detailed discussion of the  $Ra$-dependence of our findings is then provided in \S\ref{sec:Ra}.
\subsection{Spectral distribution of energy and coherence of temperature and vertical velocity } \label{sec:coh}
\begin{figure}
\begin{center}
{\includegraphics[width = \textwidth]{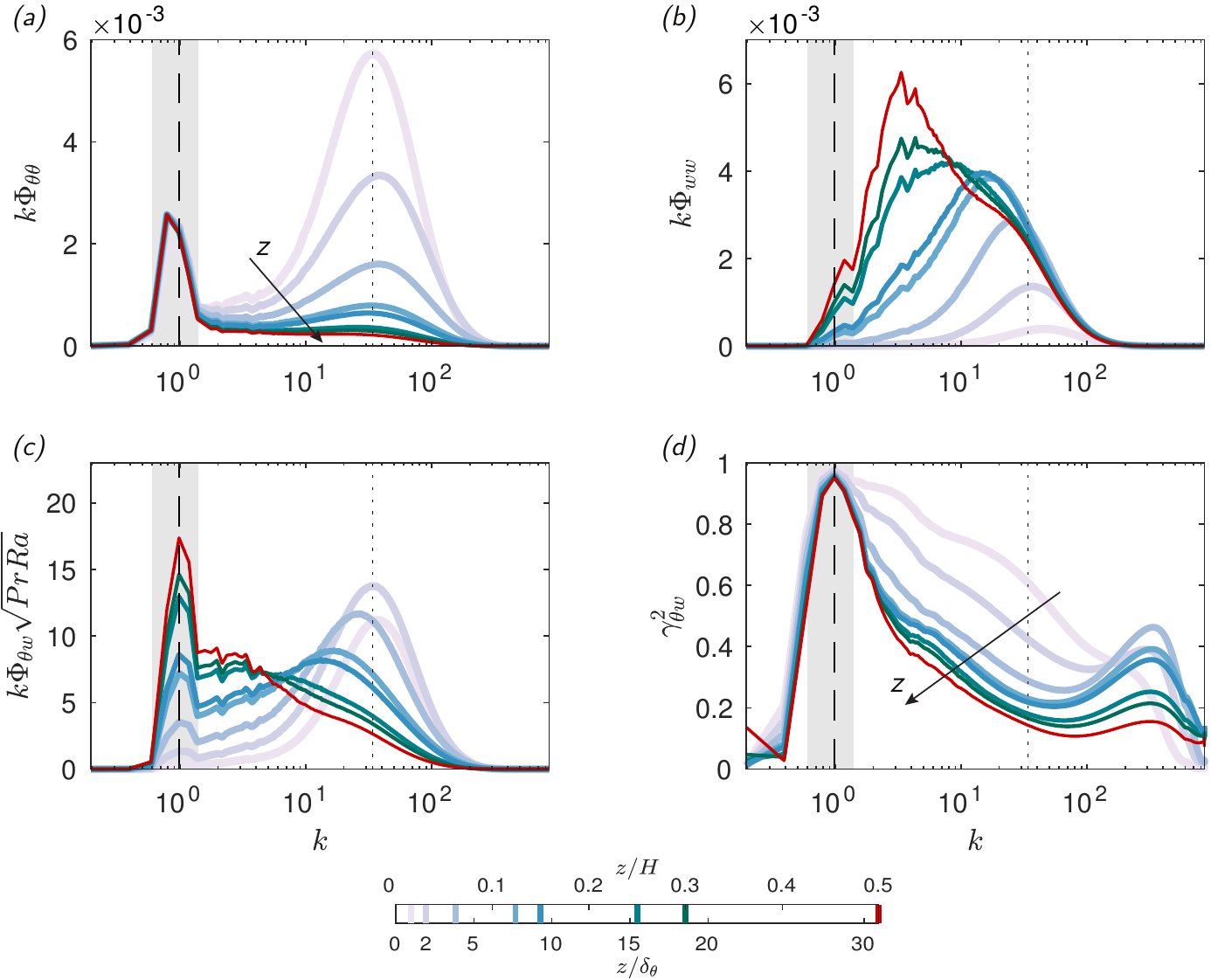}}
\caption{Premultiplied temperature (a) and vertical velocity (b) power spectra. The premultiplied co-spectrum $k\Phi_{\theta w}$ (c) is normalized such that it integrates to the turbulent heat flux. (d) Linear coherence spectrum $\gamma^2_{\theta w}$, see eq.\ (\ref{equation_coherencespectrum}). The dashed and dotted vertical lines indicate $k =1$ and $k =34$, respectively. The grey-shaded area marks the approximate range of superstructure scales $k =1 \pm0.4$. The results presented here are computed for $Ra = 10^8$. The color of the curves indicates the wall distance according to the legend below the figures.}
\label{fig:spec}
\end{center}
\end{figure}

\begin{figure}
\begin{center}
{\includegraphics[scale=.8]{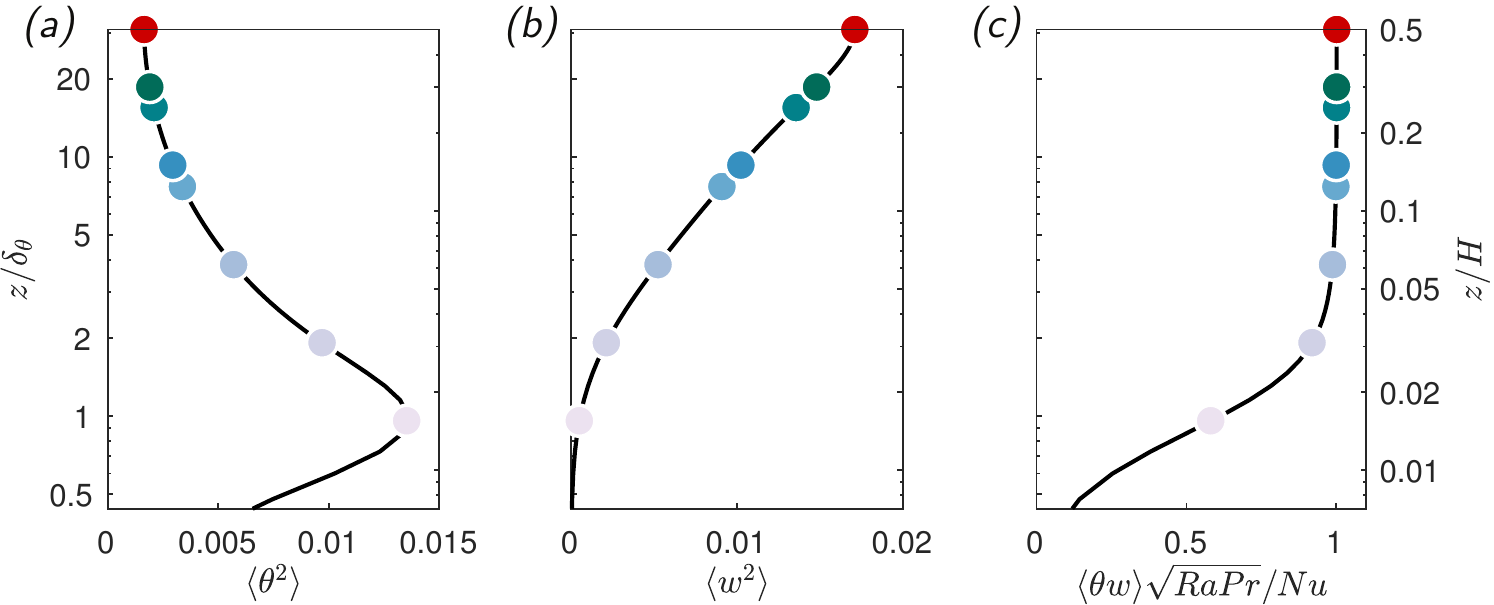}}
\caption{Wall-normal temperature (a) and vertical velocity (b) variance profiles for $Ra=10^8$. Panel (c) shows the corresponding normalized turbulent heat flux. Symbols denote the location of the spectra plotted in figures \ref{fig:spec} with corresponding colors.}
\label{fig:prof}
\end{center}
\end{figure}

To evaluate the energy distribution across different scales, we first consider the one-sided power spectra $\Phi_{\psi \psi}(k)$, where $\psi$ is a zero-mean quantity (velocity or temperature here) and $k$ is the radial wavenumber $k = \sqrt{k_x^2 +k_y^2}$. The spectra are computed for horizontal planes and averaged in time. Results for $\theta$ and $w$ at several distances from the wall are presented in figures \ref{fig:spec}a and \ref{fig:spec}b, respectively. Data are presented in  premultiplied form $k\Phi_{\psi \psi}$, such that the area under the curve equals the variance when plotted on a logarithmic scale, according to
\begin{equation}
\langle \psi^2 \rangle = \int_0^\infty \Phi_{\psi \psi} \mathrm{d}k = \int_0 ^\infty k\Phi_{\psi\psi} \mathrm{d}(\log k).
\end{equation}
For reference, the wall-normal temperature and vertical velocity variance profiles are presented in figures \ref{fig:prof}a and \ref{fig:prof}b, respectively. The symbols in these figures mark the positions at which the spectra in figure \ref{fig:spec} are computed. 

First, we focus on the situation at mid-height ($z/H = 0.5$), which corresponds to the location of the snapshots shown in figure \ref{fig:snaps}. These results are represented by the red lines in figure \ref{fig:spec}. Figure \ref{fig:spec}a shows that the temperature spectrum $k\Phi_{\theta \theta}(z/H = 0.5)$ has a pronounced peak in the range $k \approx 1 \pm0.4$ that corresponds to the superstructures \blue{(marked by a grey band as a visual aid in the figure)}. This peak contains about half of the temperature variance at $z/H = 0.5$, while the remainder of the variance is spread out over a wide range of intermediate and small-scales, which individually carry  relatively little energy. 
Figure \ref{fig:spec}b reveals that the corresponding vertical velocity spectrum $k\Phi_{w w}(z/H = 0.5)$ spans approximately the same range of scales as its temperature counterpart overall. However, its shape is significantly different as it is much more broadband and has a fairly wide peak centered around $k \approx 3.5$.  It is important to note, though, that there is significant energy in the $k\Phi_{ww}$-spectrum at the scales corresponding to the thermal superstructures, which are marked by grey shading in all panels of figure \ref{fig:spec}. This implies that velocity structures of the same size as the temperature superstructures  indeed exist. Yet, their contribution is overshadowed by stronger velocity fluctuations at smaller scales.

More insight into the correlation between the velocity and temperature structures is obtained by analyzing the one-sided co-spectrum $\Phi_{\theta w} = \Real(\langle \mathcal{F}(\theta)\mathcal{F}(w)^*\rangle)$ where $\mathcal{F}(\cdot)$ indicates the Fourier transform in the horizontal plane and $(\cdot)^*$ the complex conjugate. Figure \ref{fig:spec}c shows that the temperature-velocity co-spectrum $k\Phi_{\theta w}$ at mid-height features a pronounced large-scale peak at $k\approx 1$. This indicates that a correlation exists between the large-scale structures in $\theta$ and $w$. 
Further, $k\Phi_{\theta w}(z/H = 0.5)$  decreases with increasing $k$, but scales smaller than the superstructure size nevertheless contribute significantly to the turbulent heat transport. Aside from the degree of correlation between $\theta$ and $w$ also their magnitudes factor into the co-spectrum at a given scale. In order to  focus on the correlation aspect only, we analyze the linear coherence spectrum
\begin{equation} \label{equation_coherencespectrum}
\gamma^2_{\theta w}(k) = \frac{|\Phi_{\theta w}(k)|^2}{\Phi_{\theta \theta}(k) \Phi_{ww}(k)}.
\end{equation}
By definition, $0 \leq \gamma_{\theta w}^2 \leq 1$ and the coherence may be interpreted as the square of a scale-dependent correlation coefficient. From figure \ref{fig:spec}d it is evident that $\gamma^2_{\theta w}\geq 0.8$ for almost the entire large-scale peak with a maximum value of $\gamma^2_{\theta w} =0.95 $ at $k = 1$  \blue{(marked by a dashed line in all panels of figure \ref{fig:spec})}. The coherence quickly drops below $\gamma^2_{\theta w} =0.5 $ for larger $k$. This explains why the overall correlation coefficient between $\theta$ and $w$, which is essentially an average over the coherence spectrum, is smaller than $0.5$ as reported in \citet{stevens2018}.

\begin{figure}
\begin{center}
{\includegraphics[width = \textwidth, trim={0.5cm 0.5cm 0.7cm 0},clip]{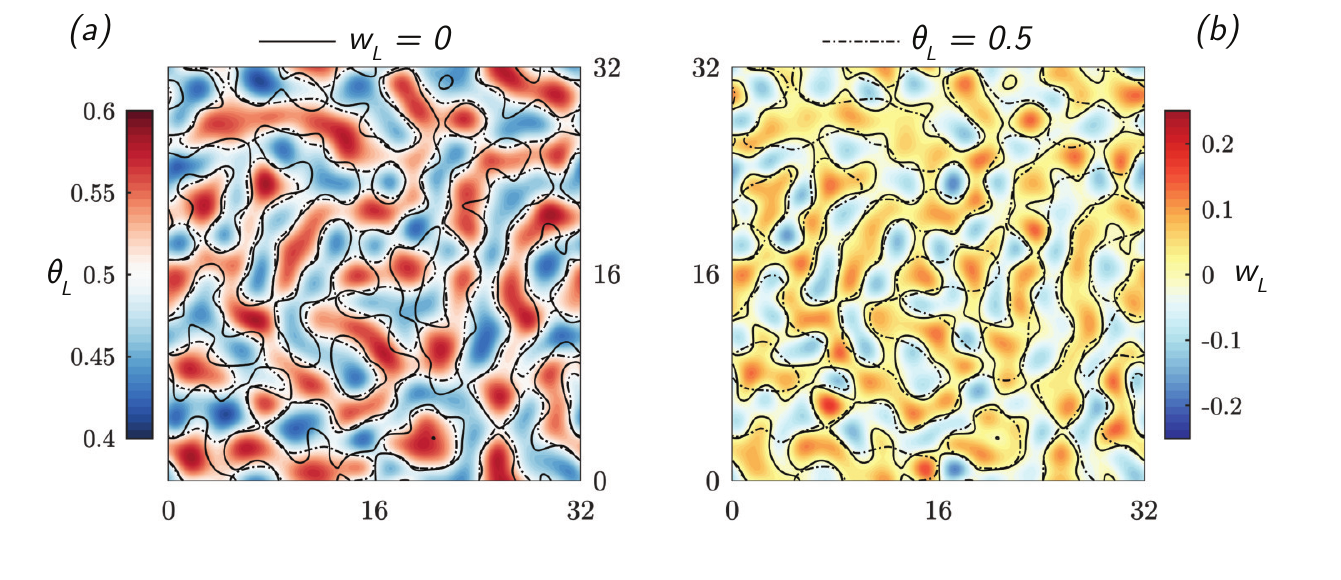}}
\caption{Same snapshots of temperature (a) and vertical velocity (b) at mid-height as presented in figure \ref{fig:snaps}, but this time filtered with a spectral low-pass filter with cut-off wavenumber $k_{\textrm{cut}}=2.5$. }
\label{fig:snapflt}
\end{center}
\end{figure}

In order to demonstrate also visually how well the large scales of $w$ and $\theta$ are correlated, we present the snapshots from figure \ref{fig:snaps} again in figure \ref{fig:snapflt}, but this time with the small-scale contributions removed. More specifically, we obtain the large-scale fields $\theta_L$ and $w_L$ using a spectral low-pass filter where the cut-off wavenumber $k_{\textrm{cut}} = 2.5$ is chosen based on the scale at which $\gamma^2_{\theta w}(z/H= 0.5)$  drops below 0.5.  Figure \ref{fig:snapflt} convincingly shows  that there is indeed a very good correspondence between patterns at the superstructure scale in temperature and vertical velocity fields, not only in size but also in location.

To summarize, we have shown that patterns corresponding to the thermal superstructures also exist in the vertical velocity.
For the vertical velocity, though,  the contribution of the superstructures in the $k\Phi_{ww}$-spectrum is sub-dominant in the sense that it does not result in a spectral peak. This has previously led to the \blue{notion} that the \textit{superstructures} in the velocity field are smaller than in the temperature field, whereas it is really the size of the \textit{most energetic structures}, as measured by the spectral peak, that is different. We will revisit the reasons for the different spectral distributions of $\theta$ and $w$ in \S \ref{sec:32}, but we first discuss the height dependence of the trends discussed so far.

Apart from the results at mid-height, figure \ref{fig:spec} also contains data at seven different wall-normal locations that span the full domain down to the thermal boundary layer thickness $\delta_\theta = 1/(2Nu)$. Remarkably, curves at all $z$-positions collapse around the peak at $k = 1$ for the temperature spectra in figure \ref{fig:spec}a. This suggests that there is very little evolution of the large-scale thermal structures along the vertical direction. Similarly, also the coherence between $\theta$ and $w$ (figure \ref{fig:spec}d) is almost independent of $z$ at the largest scales. In contrast, there is a pronounced increase in $k\Phi_{ww}$ around $k\approx 1$ with increasing distance away from the wall ---   a natural consequence of the impermeability condition at the wall. It is this increase in $k\Phi_{ww}$ that also drives a growth of the large-scale peak of the co-spectrum as $z$ increases, as shown in figure \ref{fig:spec}c.

What is striking about the $k \Phi_{\theta \theta}$ spectra  (figure \ref{fig:spec}a) is that at heights of the order of  $\delta_\theta$ there exists a second strong peak in addition to the one caused by the superstructures. 
This small-scale peak  is located at $k \approx 34$ (indicated by the dotted lines in figure \ref{fig:spec}),  which corresponds to a typical small-scale structure size of about $11 \delta_\theta$. 
Upon comparison with  figure \ref{fig:prof}a, it becomes clear that this peak carries the energy that leads to the maximum of $\langle \theta^2 \rangle$ at $z = \delta_\theta$. A similar small-scale peak is also observed for $k\Phi_{ww}$ in figure \ref{fig:spec}b, even though it is located at slightly larger $k$ in this case. For $k\Phi_{ww}$, this peak broadens towards intermediate scales with increasing $z$ and the increase of $\langle w^2 \rangle$ with increasing $z$ (see figure \ref{fig:prof}b) is mostly associated with increasing energy content at intermediate scales $k \approx 10$. 
It is further interesting to note that the spectral decomposition of $k\Phi_{\theta w}$ shifts from small-scale dominated ($z\lessapprox 3\delta_\theta$) over broadband ($0.1H\lessapprox z\lessapprox 0.2H$) to  a maximum at large scales for $z\gtrapprox 0.2H$. At the same time, the overall heat transport $\langle \theta w \rangle$ stays approximately constant beyond $z \approx 2 \delta_\theta$ (see figure \ref{fig:prof}c). In connection, these observations appear consistent with the concept of merging plumes. This was  advocated by e.g. \citet{Parodi2004},  who found that the structure size increases going away from the wall while the flux remains constant.

\subsection{Production of temperature and vertical-velocity fluctuations} \label{sec:32}

\begin{figure}
\begin{center}
{\includegraphics[scale=1.05]{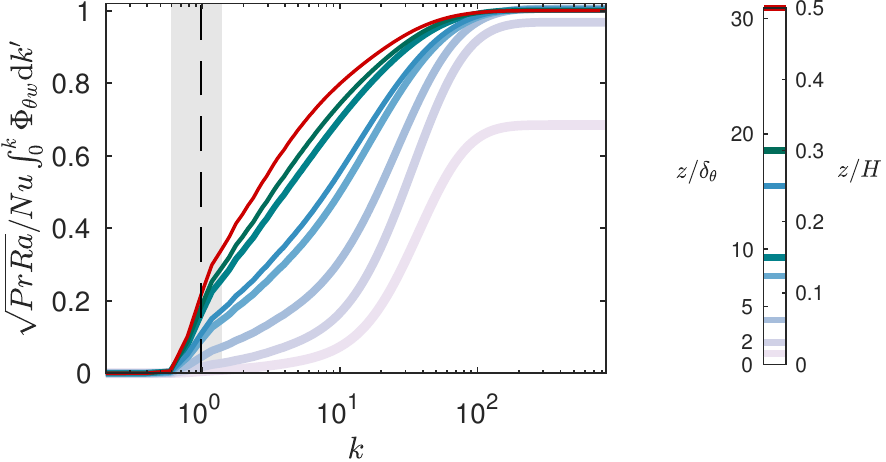}}
\caption{Cumulative co-spectrum $\int_0^k \Phi_{\theta w} \textrm{d}k^\prime$ for $Ra = 10^8$. The corresponding  co-spectra $k\Phi_{\theta w}$ are shown in figure \ref{fig:spec}c. \blue{The normalization is chosen such that the spectra add up to the relative contribution of turbulent transport to $Nu$ at each wall-height. }}
\label{fig:cum}
\end{center}
\end{figure}

In order to uncover the origin of the different spectral distribution of temperature and vertical velocity that became apparent in figure \ref{fig:spec}a,b, we now study the variance production terms of the respective variance budgets. These production terms are \citep{Deardorff1967,Kerr2001,Togni2015}
\begin{equation}
 S_\theta = -2\langle \theta w \rangle\frac{\textrm{d}\Theta}{\textrm{d}z}
 \label{eq:prodT}
\end{equation}
for $\langle \theta^2 \rangle$ and 
\begin{equation}
S_w = \langle \theta w \rangle
\label{eq:prodW}
\end{equation}
for $\langle w^2\rangle$. A trivial but nevertheless important implication that arises from comparing (\ref{eq:prodT}) and (\ref{eq:prodW}) is that $\langle \theta w \rangle$ generates $w$-variance \emph{directly}, while temperature variance is only produced in the presence of a mean gradient $\textrm{d}\Theta/\textrm{d}z$. Consequently, $S_\theta>0$ is restricted to the thermal boundary layer ($z\lessapprox \delta_\theta$) since a significant mean temperature gradient exists only there. This close to the wall $\langle \theta w \rangle$ is predominantly a small-scale quantity as evidenced by $k\Phi_{\theta w}(z = \delta_\theta)$ in figure \ref{fig:spec}c, such that $S_\theta$ is localized not only in space but also in scale. On the contrary, $\langle \theta w \rangle$ is almost independent of $z$ outside of the thermal boundary layer, see figure \ref{fig:prof}c. Hence, also $S_w$ is widely distributed across the bulk of the flow. To better understand the spectral distribution of $S_w$, we present the data from figure \ref{fig:spec}c in cumulative form in figure \ref{fig:cum}.
This figure reveals that even at $z/H=0.5$ the large-scale peak of $\Phi_{\theta w}$ only contributes about 30\% of the total flux $\langle \theta w \rangle$. In the central region of the flow the bulk of the $\langle w^2 \rangle$ production occurs at intermediate scales (say $2\lessapprox k \lessapprox 10$). This coincides with the scales at which $k\Phi_{ww}$ peaks at these wall distances, see figure \ref{fig:spec}b. This further explains why the superstructure contribution is not reflected as a spectral peak in the $k\Phi_{ww}$ spectrum. 
 
While the analysis of the production terms provides essential insight into the reasons for the different spectral decomposition of $\langle \theta^2 \rangle$ and $\langle w^2 \rangle$, other aspects cannot be addressed on this basis alone. Specifically, understanding the apparently very efficient organization of small-scale temperature fluctuations into thermal superstructures requires the analysis of inter-scale energy transfer. \blue{ Such an undertaking is beyond the scope of the present work. We note, however, that  an inverse (i.e. from  smaller to larger scales) energy transfer, is indeed observed in certain regions of the flow for both velocity and temperature when horizontally averaged budgets are considered \citep[see e.g.][]{Togni2015,Green2019} .}
 
\subsection{Wall-normal coherence of superstructures} \label{sec:cohspat}
\begin{figure}
\begin{center}
{\includegraphics[scale = 0.9]{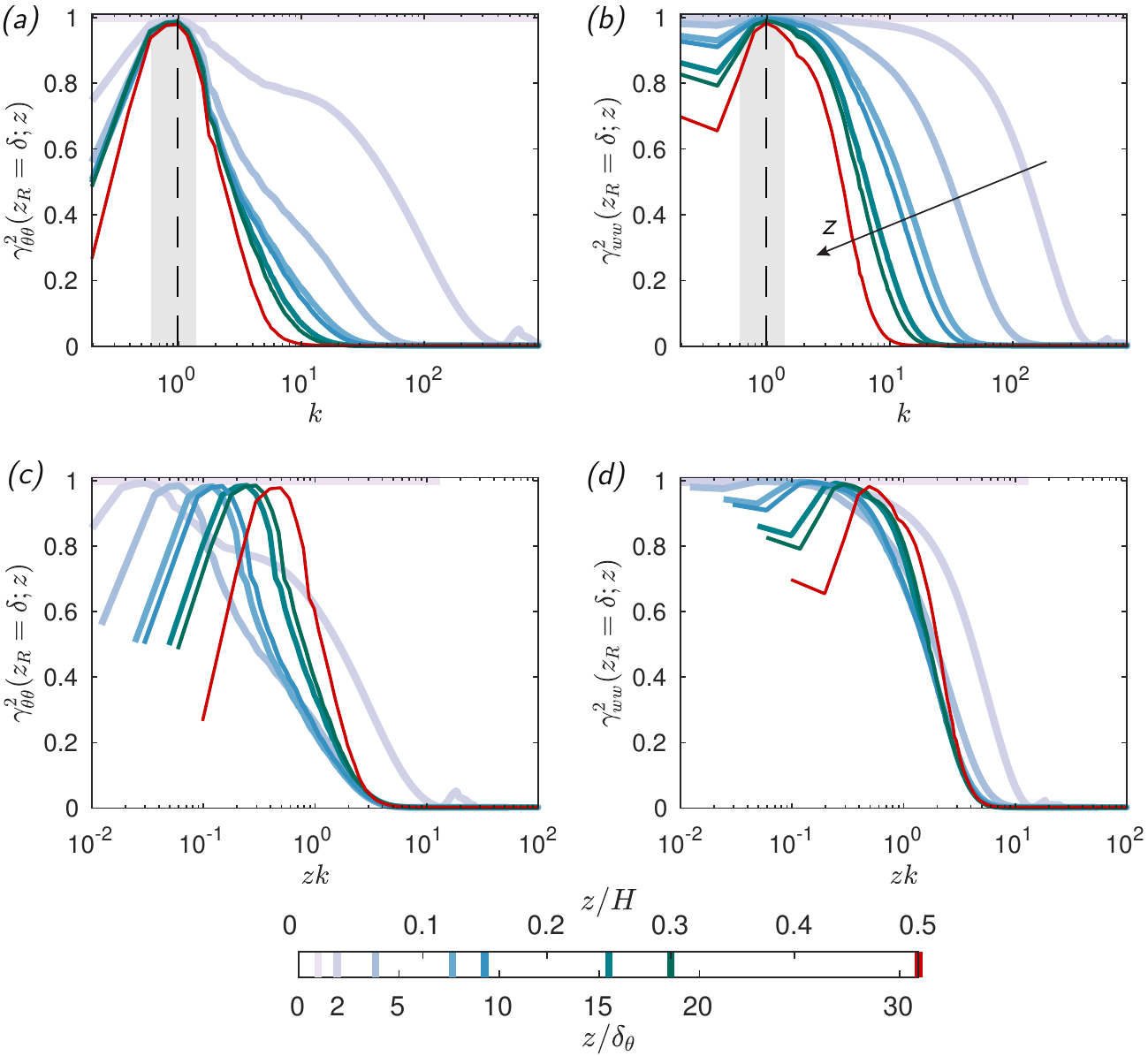}}
\caption{Spatial coherence spectra of  temperature (a) and  vertical velocity (b) with the reference plane at $z_R = \delta_\theta$. The data from (a) and (b) is plotted again in (c) and (d), respectively, as a function of $zk$ instead of $k$. All results shown are for $Ra = 10^8$. }
\label{fig:spat}
\end{center}
\end{figure}

\begin{figure}
\begin{center}
{\includegraphics[scale = 0.9]{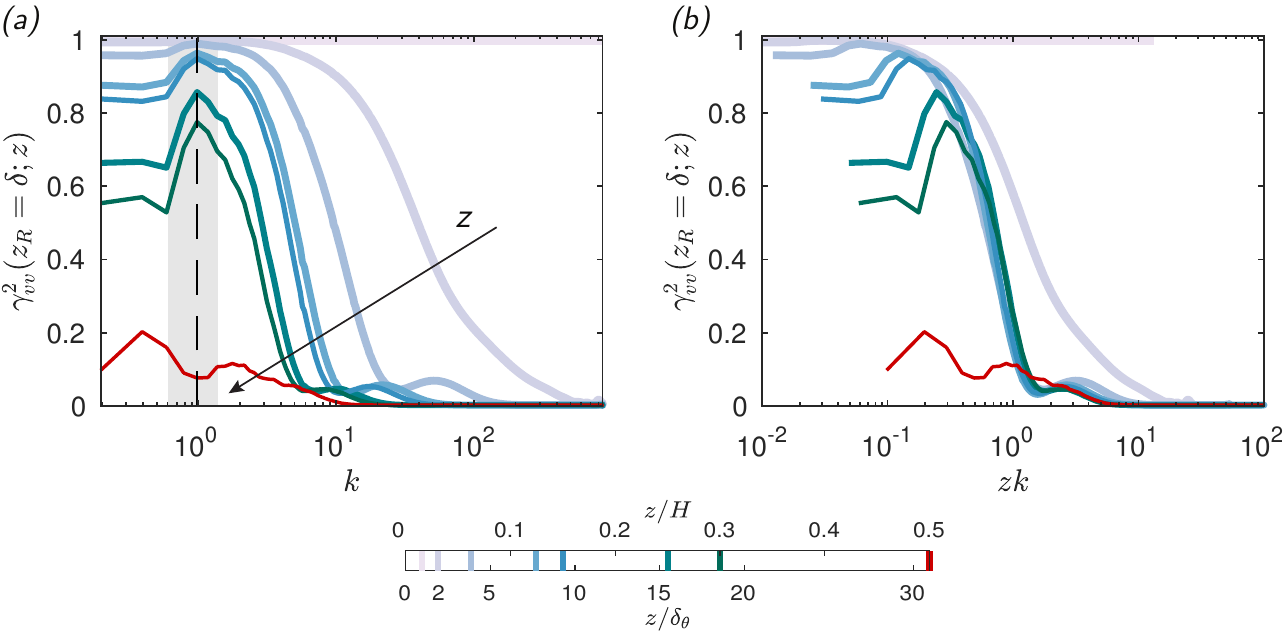}}
\caption{Spatial coherence spectra of horizontal velocity $v$  as a function of (a) $k$ and (b) $zk$.  The reference height is $z_R = \delta_\theta$ and $Ra = 10^8$.}
\label{fig:spat_hor}
\end{center}
\end{figure}

So far, we have only considered the correlation between vertical velocity and temperature at a given wall-normal location. Another important aspect is the wall-normal coherence of superstructures. \blue{There exists \blue{qualitative} evidence from comparing snapshots at different heights \citep{stevens2018} all the way down to the skin friction field \citep{Pandey2018} that an imprint of the large-scale structures is visible in the boundary layers.} To corroborate these findings in a more systematic and quantitative manner, we again turn to the linear coherence spectrum. However, this time we do not evaluate coherence between different fields, but now we evaluate the same fields at different heights $z$ and $z_R$ according to
\begin{equation}
\gamma^2_{\psi \psi}(z_R;z,k) = \frac{|\langle \mathcal{F}( \psi(z_R)) \mathcal{F}(\psi(z))^* \rangle|^2}
{\Phi_{\psi \psi}(z_R,k) \Phi_{\psi \psi}(z;k)}.
\end{equation}
We fix the reference height at $z_R = \delta_\theta$. Consequently, $\gamma^2_{\psi \psi}(z_R = \delta_\theta;z,k)$ is a measure of how correlated structures in field $\psi$ at scale $k$ and height $z$ are with fluctuations of the same scale at boundary layer height of the same field. Results are presented for temperature and vertical velocity in figures \ref{fig:spat}a and \ref{fig:spat}b, respectively. 
By definition, the result at $z = \delta_\theta$ is the correlation with the reference itself and therefore $ \gamma^2_{\psi \psi}(z_R = \delta_\theta;\delta_\theta,k)=1$ trivially. In the  superstructure peak ($k =1 \pm 0.4$, marked by grey shading in the figure) $ \gamma^2_{\psi \psi}(z_R = \delta_\theta;z,k)$ is close to one, even at mid-height. This holds for both temperature and vertical velocity and  implies a very strong degree of spatial coherence for the largest structures in both fields. 
Differences between $\theta$ and $w$ only occur at smaller scales. Beyond $z = 2 \delta_\theta$ the spatial coherence of $\theta$ decreases very quickly as a function of $k$, but has a limited $z$-dependence. In contrast, curves for $\gamma^2_{ww}$ in figure \ref{fig:spat}b show significant variation with $z$ with the decline occurring at progressively smaller $k$ with increasing $z$.
Apart from quantifying the correlation, $\gamma^2_{\psi \psi}(z_R= \delta_\theta)$ also provides information about the self-similarity of structures that are connected or `attached' to the thermal boundary layer. This is of interest since previous authors \citep[e.g.][]{He2014, Ahlers2014} have referred to the attached-eddy framework \citep{Townsend1976,Perry1982,Marusic2019}, which assumes the existence of self-similar wall-attached structures, in the interpretation of their results. For the coherence spectrum, self-similarity implies that curves of $\gamma^2_{\psi \psi}(z_R= \delta_\theta)$ should collapse if plotted against $zk$, that is if the scale is normalized by the distance from the wall \citep[see][]{Baars2017, Krug2019}. 
We test this for temperature and vertical velocity in figures \ref{fig:spat}c,d. Clearly, self-similarity is not observed for the temperature (figure \ref{fig:spat}c). However, figure \ref{fig:spat}d shows that the data for $w$ indeed collapse to a reasonable degree for $3\delta_\theta \lessapprox z  \lessapprox 0.3H$. 
To check if self-similarity scaling in this range is a property of the velocity field in generally, we additionally present results for $\gamma^2_{vv}(z_R= \delta_\theta)$, where $v$ is the horizontal velocity component, in figure \ref{fig:spat_hor}. The vertical coherence of $v$ also exhibits the same superstructure peak as observed for the other quantities, which can be seen in figure \ref{fig:spat_hor}a. Only its magnitude decreases with increasing $z$ and is close to zero at mid-height. This is consistent with the roll structures not having a horizontal component at $z/H \approx 0.5$ and also the spectral energy $\Phi_{vv}(k =1\pm0.4)$  (not shown) is minimal there. As figure \ref{fig:spat_hor}b shows, $\gamma^2_{vv}(z_R= \delta_\theta)$ displays the same collapse when plotted versus $zk$ and in the same range of $z$ as previously observed for $w$. This means that for the velocity fields in a significant part of the domain (at least $3 \delta_\theta \lessapprox z \lessapprox 0.3H$) structures attached to the boundary layer display self-similar behavior. The same trends are observed at different $Ra$ but are not shown here for brevity. 

\subsection{Rayleigh number trends} \label{sec:Ra}
\begin{figure}
\begin{center}
{\includegraphics[scale=.8]{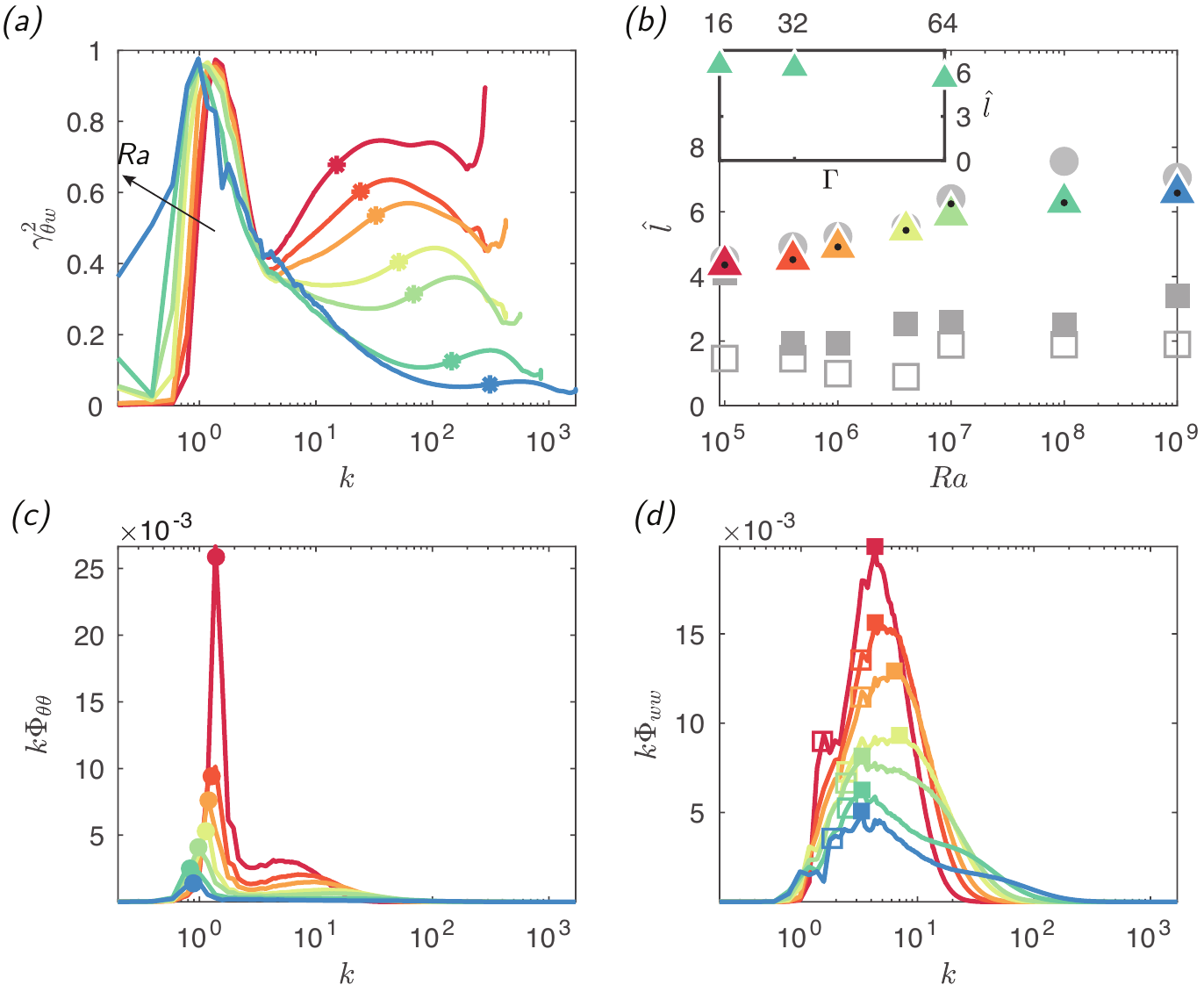}}
\caption{(a) Coherence spectrum at mid-height for $10^5\leq Ra \leq 10^9$; see panel (b) for the color-code. Stars indicate the wavenumber corresponding to $10\eta$ at the respective $Ra$. (b) Wavelength $\hat{l}$ of the spectral peaks of $\gamma^2_{\theta w}$ (triangles), $\Phi_{\theta \theta}$ (circles), $\Phi_{ww}$ (filled squares), $k\Phi_{ww}$ (open squares) \blue{ and $k\Phi_{\theta w}$ (black dots)}. The corresponding spectra $k\Phi_{\theta \theta}$ and $k\Phi_{ww}$ are shown in panels (c) and (d) with symbols marking the peak locations as described for (b). \blue{The inset in (b) additionally shows the aspect ratio dependence of $\hat{l}$ based on $\gamma^2_{\theta w}$  at $Ra = 10^8$, see appendix \ref{app} for details.}}
\label{fig:Racoh}
\end{center}
\end{figure}

As a final point, we study the $Ra$ number dependence of the properties discussed in \S\ref{sec:coh}. To this end, we present results for $\gamma^2_{\theta w}$ evaluated at mid-height for $10^5\leq Ra \leq 10^9$ in figure \ref{fig:Racoh}a. 
The magnitude and the shape of the large-scale peak are nearly independent of $Ra$. However, the peak location shifts towards smaller $k$ with increasing $Ra$. The corresponding increase in the large-scale structure is quantified in figure \ref{fig:Racoh}b, where the triangles indicate the structure size ($\hat{l} = 2\pi /\hat{k}$) corresponding to the peak in the coherence. Here, the peak location $\hat{k}$ is obtained from fitting a parabola to three points centered around the peak of $\gamma^2_{\theta w}$ and the results are also listed in table \ref{table1}. Evidently, $\hat{l}$ is significantly larger than the wavelength of the structure at the onset of convection, which is $\approx 2$ \citep[e.g.][]{Drazin2004}.  Additionally, length scales corresponding to the spectral peak in $\Phi_{\theta \theta}$ (circles), $\Phi_{ww}$ (filled squares), and $k\Phi_{ww}$ (open squares) are included in figure \ref{fig:Racoh}b and the corresponding spectra are shown in panels (c,d) of the same figure. The spectral peak from the temperature spectrum corresponds to slightly larger length scales compared to the results based on $\gamma^2_{\theta w}$, but the differences are quite small. 
Due to its broadband nature, the spectral peaks for the vertical velocity are found at a different location in the regular and premultiplied spectra. The maximum of $\Phi_{ww}$ only agrees with the results based on coherence and temperature at $Ra = 10^5$. For higher $Ra$ fluctuations at intermediate length scales dominate the velocity spectrum. Therefore,  the use of the velocity spectra leads to significantly lower estimates for $\hat{l}$ than the temperature spectra at higher $Ra$, as mentioned in \S\ref{sec:coh}. 

\blue{The dependence of the superstructure size on the aspect ratio $\Gamma$ of the periodic domain was already discussed in \citet{stevens2018}. From their results, it appears that the superstructure size based on the peak in $\Phi_{\theta \theta}$ increases monotonically (albeit slowly for $\Gamma>16$) with increasing $\Gamma$. The inset of figure \ref{fig:Racoh}b shows that $\hat{l}$ based on $\gamma^2_{\theta w}$ decreases slightly if $\Gamma$ is increased from 16 to 64. This difference is not rooted in the fact that a different metric is employed here, but is caused by an error in the computation of the spectra presented in \citet{stevens2018}. We plot the recomputed spectra in appendix \ref{app} and these show that the temperature spectral peak indeed exhibits the same trend.}

The concept of a convection roll, i.e. a thermally driven velocity structure, suggests  to define the superstructure size in RBC as the scale where the correlation between temperature and velocity fields is maximum.  We therefore argue that conceptually the most straightforward way to define the superstructure size is via the coherence spectrum. It should be noted that the coherence peak is not necessarily coincident with the peak of  the co-spectrum due to the different distributions of $\Phi_{\theta \theta}$ and  $\Phi_{ww}$. In practice, however,  the peaks of $\gamma^2_{\theta w}$ and $k\Phi_{\theta w}$ coincide within measurement accuracy for the cases presented here \blue{(see figure \ref{fig:Racoh}b)}. \blue{This seems to be a consequence of the sharp drop-off of $\gamma^2_{\theta w}$ and $\Phi_{\theta \theta}$ with  increasing $k$ that outweighs the increase in $\Phi_{ww}$. The situation may change however, e.g. for different $Pr$ numbers.  Some caution is therefore advised in this matter. For a case in point, we note that the small-scale peak in  $k\Phi_{\theta w}$ at $z = \delta_\theta$ (figure \ref{fig:spec}c) is without counterpart in $\gamma^2_{\theta w}$ (figure \ref{fig:spec}c). This  indicates that the peak in turbulent transport is predominantly driven by magnitude, not coherence. }  \blue{The peaks of the $\Phi_{ww}$ and the $k\Phi_{ww}$ spectra may be misleading  as indicators for superstructure size since the velocity spectra are dominated by motions at intermediate length scales.}

As an aside, we discuss the increase of $\gamma^2_{\theta w}$ that is seen to occur at high $k$ in figure \ref{fig:Racoh}a. This increase at small-scales occurs for lower values of $k$ and is stronger for the lower $Ra$. A comparison with figure \ref{fig:Racoh}c,d reveals that there is only minimal energy at these small scales. These observations are consistent with the notion that the higher values of $\gamma^2_{\theta w}$ mark the transition to a viscous dominated regime. In the viscous regime, the correlation between $\theta$ and $w$ is high because the balance is predominantly  between buoyancy and viscous forces. \blue{This is similar to the situation at the onset of convection, where the correlation between velocity and temperature fluctuations is very high \citep{Bodenschatz2000}.}
To lend support to this understanding, we added the length scale $10\eta$ as a reference scale for the viscous regime in the figure. Here, $\eta = (\nu^3/\langle \varepsilon\rangle_V)^{1/4}$ is the Kolmogorov length-scale and $\langle \varepsilon \rangle_V$ is the volume-averaged dissipation rate obtained from the identity $\langle \varepsilon \rangle_V = (Nu -1)/\sqrt{Ra Pr}$. \blue{It is seen  in figure \ref{fig:Racoh}a that the scale at which the high-wavenumber increase of $\gamma^2_{\theta w}$ occurs roughly coincides with $10\eta$ for $Ra>10^6$, just as  expected from the above. The agreement is less good for the (marginally turbulent) cases at even lower $Ra$, where the increase in $\gamma^2_{\theta w}$ starts at scales significantly larger than $10\eta$.  }

\begin{figure}
\begin{center}
{\includegraphics[scale=.8]{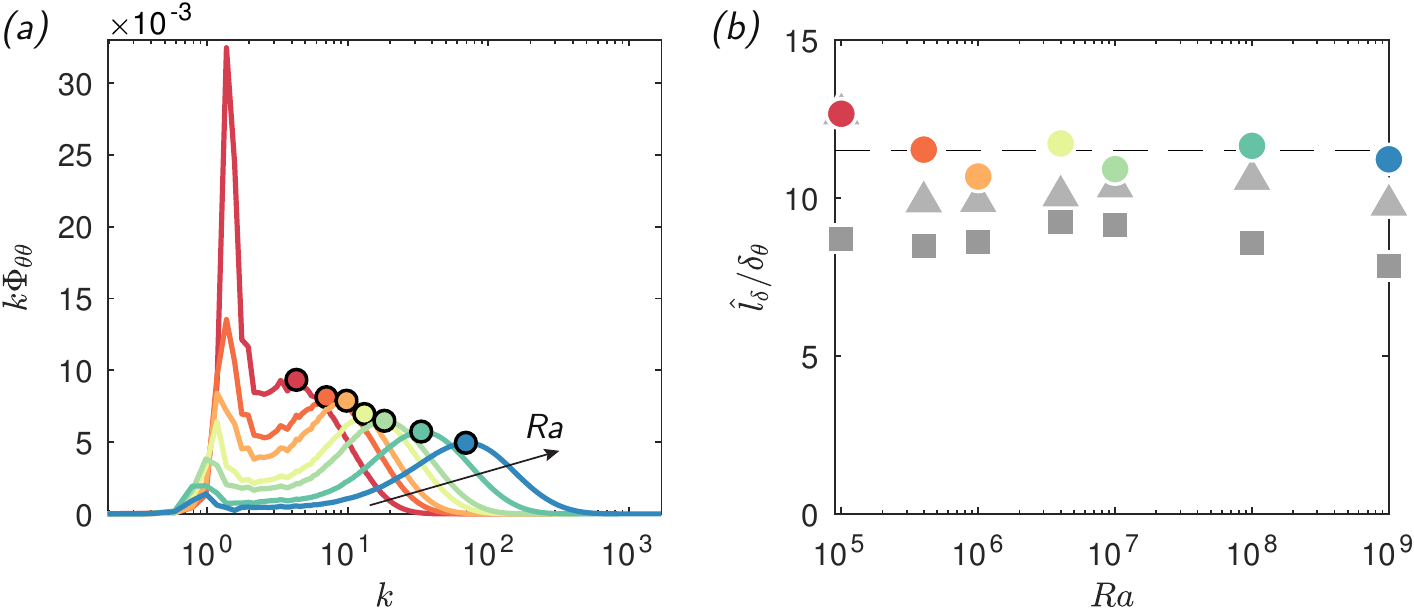}}
\caption{(a) Premultiplied temperature power spectra $k\Phi_{\theta \theta}$ at $z = \delta_\theta$ for $10^5\leq Ra \leq 10^9$. Symbols mark the location of the small-scale peak determined as the maximum of $k\Phi_{\theta \theta}(k)$ for $k>2$. Note that at the lower $Ra$, this peak does not correspond to a global maximum of $k\Phi_{\theta \theta}$. (b) Length scale $ \hat{l}_\delta$ associated with the small-scale peak of $k\Phi_{\theta \theta}$ (circles), $k\Phi_{ww}$ (squares) and $k\Phi_{\theta w}$ (triangles) normalized with the thermal boundary layer thickness $\delta_\theta$. The dashed line is at 11.5 for reference. }
\label{fig:Rasmall}
\end{center}
\end{figure}

The $Ra$ dependence of the near-wall characteristics of the temperature field are displayed in figure \ref{fig:Rasmall} in which $k\Phi_{\theta \theta}$ is plotted at $z = \delta_\theta$ for each $Ra$. This figure shows that the small-scale peak contributes an increasingly larger part of the total energy with increasing $Ra$. 
At the same time, the scale separation between the small-scales and the large-scale superstructures increases with increasing thermal driving. We define the length scale of the small-scale structures as $\hat{l}_\delta = 2 \pi/\hat{k}_\delta$, where $\hat{k}_\delta$ is the location of the high-$k$ peak. Figure \ref{fig:Rasmall}b shows that $\hat{l}_\delta$ is approximately constant for $10^5\leq Ra \leq 10^9$, when normalized with the boundary layer thickness $\delta_\theta$. The magnitude of the ratio $\hat{l}_\delta/\delta_\theta$ differs slightly depending on the quantity considered. The most energetic small-scale structures for the temperature are about $11.5\delta_\theta$, for $w$ it is about $8.5\delta_\theta$, and $k\Phi_{\theta w}$ peaks at about $10 \delta_\theta$.

\section{Conclusion} \label{sec:conc}
Contrary \blue{to what prior analysis  \citep{stevens2018,Pandey2018} appeared to suggest}, we found  that superstructures of approximately  \emph{the same size} exist in the temperature and vertical velocity fields in large-aspect ratio Rayleigh-B\'{e}nard flow. These result in a very significant large-scale peak in the linear coherence spectrum of $\theta$ and $w$ that signifies almost perfect correlation at the large length scales. Unlike it is the case for $\theta$, we find that the superstructures in $w$ do not correspond to a spectral peak in the power spectrum of $w$. This difference has previously led to the above-mentioned confusion regarding potentially different sizes of the largest structures in $\theta$ and $w$. The fact that the most energetic motions, as measured by the peak in the spectra, occur at intermediate scales for $w$, but at the superstructure scale for $\theta$, can be explained by differences in the production terms of the respective variance budgets. In particular, temperature production is confined to the boundary layer and small-scales, while buoyancy forcing acts at intermediate scales and throughout the entire bulk of the flow. Furthermore, we find that the superstructure scale increases with $Ra$ for $10^5\leq Ra \leq 10^9$, i.e. the full range investigated here, when the structure size is based on the coherence spectrum as suggested.
 It should be noted that integral length scales of temperature and turbulent kinetic energy as used in \citet{stevens2018} do not accurately capture this growth, which shows the importance of selecting the appropriate metric to quantify superstructures in RBC.

In agreement with previous observations of superstructure footprints in the boundary layer region, we find an almost perfect spatial correlation of the superstructure scales from the boundary-layer height $\delta_\theta$ up to mid-height, for both $\theta$ and $w$. Also, the temperature spectra, as well as $\gamma^2_{\theta w}$, are seen to collapse at different heights. 
Hence, there is no noticeable dependence of the superstructure scale on $z$, effectively ruling out a significant growth of the thermal structures due to horizontal transport while they are travelling upward as was suggested by \citet{Pandey2018}. The decrease of spatial correlation (quantified by the linear coherence spectrum) at intermediate scales when increasing the distance to the reference height $\delta_\theta$ is observed to follow a self-similar trend for $w$ and the horizontal component $v$, but not for $\theta$. The reason for this difference remains unclear but warrants further investigation.

Moreover, we find that the energy distribution of the temperature field is bimodal. Besides the $z$-independent large-scale contribution of the superstructures, premultiplied spectra reveal the existence of a pronounced small-scale peak at boundary-layer height. 
The two peaks are separated by a spectral gap that increases with $Ra$, which is also visible in the co-spectra of $\theta$ and $w$. However, $k\Phi_{ww}$ displays a small-scale peak only near the wall and is broadband otherwise. For the temperature fluctuations, the small-scale peak carries the energy that leads to the maximum of $\langle \theta^2\rangle$ at $z = \delta_\theta$ \citep[see e.g.][]{Wang2016}. The length scale of the associated structures is approximately $l_\delta \approx 10 \delta_\theta$. 
It is interesting to note that the situation described here has a close resemblance to findings in turbulent boundary layers. There an `inner peak' is observed that is fixed at 15  viscous units ($l_{visc}$) away from the wall and with typical streamwise length scales of  about $1000 l_{visc}$ \citep{Hutchins2007large}. The scale separation between the `inner peak' and the large-scale structures is, however, significantly stronger in RBC. This appears to suggest distinctly different processes, as was already pointed out in \citet{Pandey2018} in a different context, and raises questions about their interaction. A better understanding of these aspects  will be very insightful for modelling approaches.

\acknowledgements{The authors acknowledge stimulating discussions with Woutijn Baars and thank Alexander Blass for help with the data. This work is supported by the Twente Max-Planck Center, the German Science Foundation (DFG) via program SSP 1881, and the  ERC (the European Research Council) Starting Grant No. 804283 UltimateRB. The authors gratefully acknowledge the Gauss Centre for Supercomputing e.V. ({\color{blue}\url{www.gauss-centre.eu}}) for funding this project by providing computing time on the GCS Supercomputer SuperMUC-NG at Leibniz Supercomputing Centre ({\color{blue}\url{www.lrz.de}}). Part of the work was carried out on the national e-infrastructure of SURFsara, a subsidiary of SURF cooperation, the collaborative ICT organization for Dutch education and research.}

Declaration of Interests. The authors report no conflict of interest.

\appendix
\section{\blue{Aspect ratio dependence of the superstructure size}} 
\label{app}
\begin{figure}
\begin{center}
{\includegraphics[scale = 0.8]{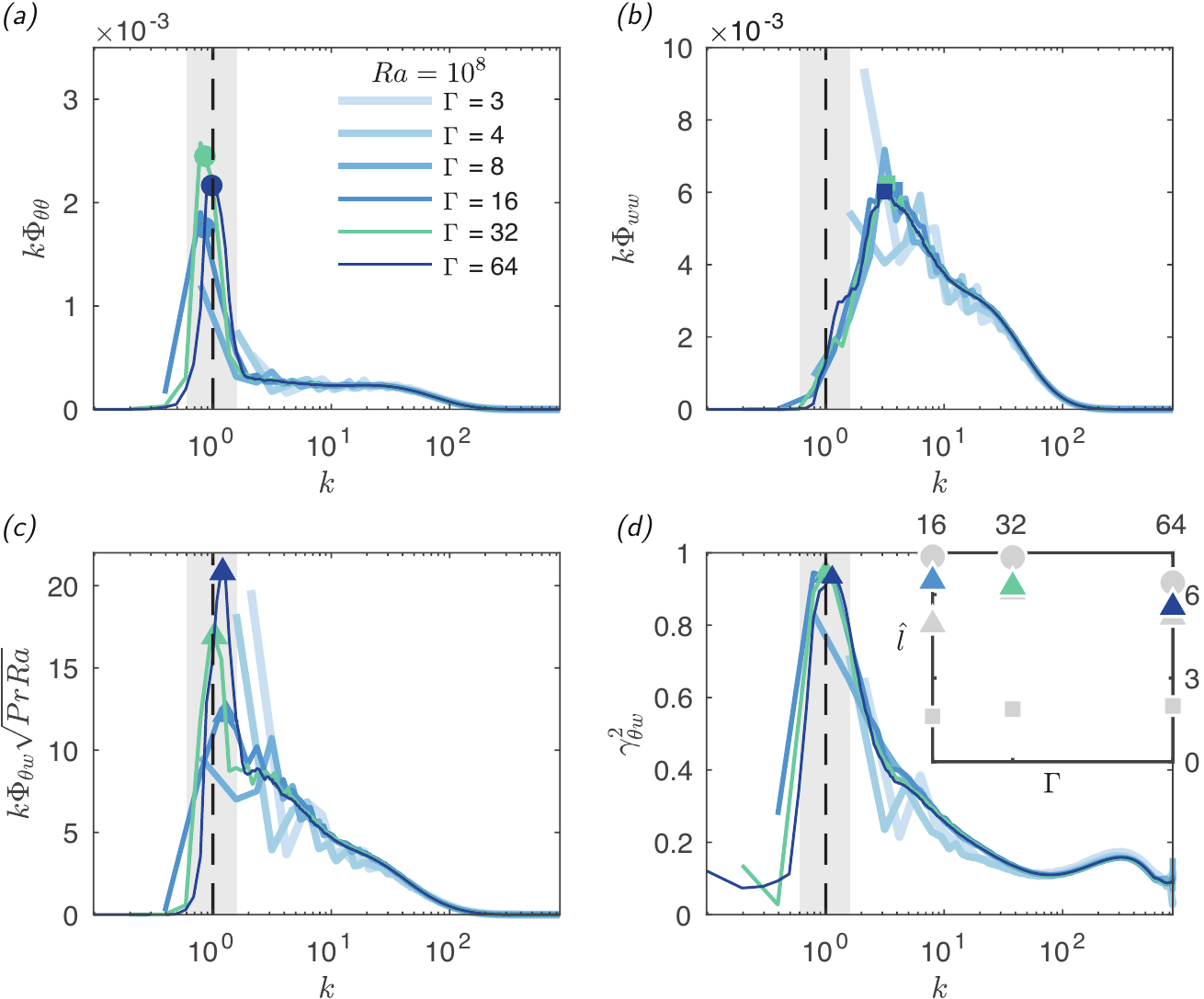}}
\caption{\blue{Premultiplied temperature (a) and vertical velocity (b) power spectra. The premultiplied co-spectrum $k\Phi_{\theta w}$ (c) is normalized such that it integrates to the turbulent heat flux. (d) Linear coherence spectrum $\gamma^2_{\theta w}$.  Symbols indicate the peak values for the three largest $\Gamma$. The inset shows $\hat{l}$ based on coherence (colored triangles) and co-spectra (grey triangles), as well as on temperature (circles) and velocity (squares) power spectra.  The dashed  vertical line indicates $k =1$ and the grey-shaded area marks the approximate range of superstructure scales $k =1 \pm0.4$ (same as in figure \ref{fig:spec}). All results presented here are computed for $Ra = 10^8$ at mid-height; the legend in (a) applies to all panels. } }
\label{fig:gam}
\end{center}
\end{figure}

\blue{The aspect ratio dependence of the superstructures has been studied in \citet{stevens2018} before. However, there was a bug in the radial averaging of the spectra, which led to a somewhat altered spectral distribution, especially at the largest scales. In figure \ref{fig:gam}, we present results  at $Ra = 10^8$ for aspects ratios varying between 3 and 64. Up to $\Gamma= 8$ there is no clear peak as the highest values are attained for the smallest $k$ for all quantities displayed. Distinct large-scale peaks emerge for $\Gamma \geq 16$ for $k\Phi_{\theta \theta}$ (figure \ref{fig:gam}a), $k\Phi_{\theta w}$ (figure \ref{fig:gam}c), and $\gamma^2_{\theta w}$ (figure \ref{fig:gam}d).  The locations of these peaks (again obtained by fitting a parabola over three points) are shown in figure \ref{fig:gam}. Results based on $k\Phi_{\theta \theta}$ and $\gamma^2_{\theta w}$ decrease monotonically with increasing $\Gamma$, while $\hat{l}$ based on the co-spectrum increases between $\Gamma = 16$ and $\Gamma = 32$. The peaks in $k \Phi_{w w}$, and remarkably these spectra  in general, exhibit only a minor sensitivity to $\Gamma$. The peaks of the velocity spectra remain, however, at intermediate scales significantly smaller than the superstructure size.  }

\blue{The conclusion of \citet{stevens2018} that very large domains are needed to fully converge the superstructure size remains valid, even though, at least for the values checked here, the trend is opposite (decreasing size) to what was previously believed. The scale $\hat{l}$ still varies by about 10\% between the cases with $\Gamma = 32$ and $\Gamma = 64$. The more basic requirement is however that $\Gamma \gtrapprox 16$ because only then the large-scale peak is actually resolved.  }

\bibliographystyle{jfm}
\bibliography{rb_struct}

\begin{thebibliography}{49}
\expandafter\ifx\csname natexlab\endcsname\relax\def\natexlab#1{#1}\fi
\def\au#1{#1} \def\ed#1{#1} \def\yr#1{#1}\def\at#1{#1}\def\jt#1{\textit{#1}}
  \def\bt#1{#1}\def\bvol#1{\textbf{#1}} \def\vol#1{#1} \def\pg#1{#1}
  \def\publ#1{#1}\def\arxiv#1{#1}\def\org#1{#1}\def\st#1{\textit{#1}}

\bibitem[Ahlers {\em et~al.\/}(2014)Ahlers, Bodenschatz \& He]{Ahlers2014}
{\sc \au{Ahlers, G.}, \au{Bodenschatz, E.} \& \au{He, X.}} \yr{2014}
  \at{{Logarithmic temperature profiles of turbulent Rayleigh--B\'{e}nard
  convection in the classical and ultimate state for a Prandtl number of 0.8}}.
   \jt{J. Fluid Mech.}  \bvol{758},  \pg{436--467}.

\bibitem[Ahlers {\em et~al.\/}(2009)Ahlers, Grossmann \& Lohse]{Ahlers2009}
{\sc \au{Ahlers, G.}, \au{Grossmann, S.} \& \au{Lohse, D.}} \yr{2009}
  \at{{Heat transfer and large scale dynamics in turbulent Rayleigh-B\'{e}nard
  convection}}.  \jt{Rev. Mod. Phys.}  \bvol{81}~(2),  \pg{503--537}.

\bibitem[Baars {\em et~al.\/}(2017)Baars, Hutchins \& Marusic]{Baars2017}
{\sc \au{Baars, W.~J.}, \au{Hutchins, N.} \& \au{Marusic, I.}} \yr{2017}
  \at{Self-similarity of wall-attached turbulence in boundary layers}.  \jt{J.
  Fluid Mech.}  \bvol{823}~(R2).

\bibitem[Bailon-Cuba {\em et~al.\/}(2010)Bailon-Cuba, Emran \&
  Schumacher]{bai10}
{\sc \au{Bailon-Cuba, J.}, \au{Emran, M.~S.} \& \au{Schumacher, J.}} \yr{2010}
  \at{Aspect ratio dependence of heat transfer and large-scale flow in
  turbulent convection}.  \jt{J. Fluid Mech.}  \bvol{655},  \pg{152--173}.

\bibitem[Bodenschatz {\em et~al.\/}(2000)Bodenschatz, Pesch \&
  Ahlers]{Bodenschatz2000}
{\sc \au{Bodenschatz, E.}, \au{Pesch, W.} \& \au{Ahlers, G.}} \yr{2000}
  \at{{Recent developments in Rayleigh-B{\'e}nard convection}}.  \jt{Annu. Rev.
  Fluid Mech.}  \bvol{32}~(1),  \pg{709--778}.

\bibitem[{Chill\`{a}} \& Schumacher(2012)]{Chilla2012}
{\sc \au{{Chill\`{a}}, F.} \& \au{Schumacher, J.}} \yr{2012}  \at{{New
  perspectives in turbulent Rayleigh-B\'{e}nard convection}}.  \jt{Eur. Phys.
  J. E}  \bvol{35}~(7),  \pg{58}.

\bibitem[Cierpka {\em et~al.\/}(2019)Cierpka, K{\"a}stner, Resagk \&
  Schumacher]{Cierpka2019}
{\sc \au{Cierpka, C.}, \au{K{\"a}stner, C.}, \au{Resagk, C.} \& \au{Schumacher,
  J.}} \yr{2019}  \at{{On the challenges for reliable measurements of
  convection in large aspect ratio Rayleigh-B{\'e}nard cells in air and
  sulfur-hexafluoride}}.  \jt{Exp. Therm. Fluid Sci.}  \pg{p. 109841}.

\bibitem[Deardorff \& Willis(1967)]{Deardorff1967}
{\sc \au{Deardorff, J.~W.} \& \au{Willis, G.~E.}} \yr{1967}  \at{Investigation
  of turbulent thermal convection between horizontal plates}.  \jt{J. Fluid
  Mech.}  \bvol{28}~(4),  \pg{675--704}.

\bibitem[Drazin \& Reid(2004)]{Drazin2004}
{\sc \au{Drazin, P.~G.} \& \au{Reid, W.~H.}} \yr{2004} {\em Hydrodynamic
  stability\/}.  \publ{Cambridge Univ. Press}.

\bibitem[Du~Puits {\em et~al.\/}(2013)Du~Puits, Resagk \& Thess]{duPuits2013}
{\sc \au{Du~Puits, R.}, \au{Resagk, C.} \& \au{Thess, A.}} \yr{2013}
  \at{{Thermal boundary layers in turbulent Rayleigh--B{\'e}nard convection at
  aspect ratios between 1 and 9}}.  \jt{New J. Phys.}  \bvol{15}~(1),
  \pg{013040}.

\bibitem[Emran \& Schumacher(2015)]{Emran2015}
{\sc \au{Emran, M.~S.} \& \au{Schumacher, J.}} \yr{2015}  \at{Large-scale mean
  patterns in turbulent convection}.  \jt{J. Fluid Mech.}  \bvol{776},
  \pg{96--108}.

\bibitem[Fitzjarrald(1976)]{Fitzjarrald1976}
{\sc \au{Fitzjarrald, D.~E.}} \yr{1976}  \at{An experimental study of turbulent
  convection in air}.  \jt{J. Fluid Mech.}  \bvol{73}~(4),  \pg{693--719}.

\bibitem[Green {\em et~al.\/}(2019)Green, Vlaykov, Mellado \&
  Wilczek]{Green2019}
{\sc \au{Green, G.}, \au{Vlaykov, D.~G.}, \au{Mellado, J.~P.} \& \au{Wilczek,
  M.}} \yr{2019}  \at{{Resolved energy budget of superstructures in
  Rayleigh-B\'{e}nard convection}}.  \jt{arXiv preprint arXiv:1905.10278} .

\bibitem[Hartlep {\em et~al.\/}(2003)Hartlep, Tilgner \& Busse]{Hartlep2003}
{\sc \au{Hartlep, T.}, \au{Tilgner, A.} \& \au{Busse, F.~H.}} \yr{2003}
  \at{{Large scale structures in Rayleigh-B{\'e}nard convection at high
  Rayleigh numbers}}.  \jt{Phys. Rev. Lett.}  \bvol{91}~(6),  \pg{064501}.

\bibitem[Hartlep {\em et~al.\/}(2005)Hartlep, Tilgner \& Busse]{har05}
{\sc \au{Hartlep, T.}, \au{Tilgner, A.} \& \au{Busse, F.~H.}} \yr{2005}
  \at{Transition to turbulent convection in a fluid layer heated from below at
  moderate aspect ratio}.  \jt{J. Fluid Mech.}  \bvol{544},  \pg{309--322}.

\bibitem[He {\em et~al.\/}(2014)He, van Gils, Bodenschatz \& Ahlers]{He2014}
{\sc \au{He, X.}, \au{van Gils, D. P.~M.}, \au{Bodenschatz, E.} \& \au{Ahlers,
  G.}} \yr{2014}  \at{{Logarithmic spatial variations and universal $f^{-1}$
  power spectra of temperature fluctuations in turbulent Rayleigh--B\'{e}nard
  convection}}.  \jt{Phys. Rev. Lett.}  \bvol{112}~(17),  \pg{174501}.

\bibitem[Hogg \& Ahlers(2013)]{hog13}
{\sc \au{Hogg, J.} \& \au{Ahlers, G.}} \yr{2013}  \at{{{Reynolds}}-number
  measurements for low-{{Prandtl}}-number turbulent convection of
  large-aspect-ratio samples}.  \jt{J. Fluid Mech.}  \bvol{725},
  \pg{664--680}.

\bibitem[Huisman {\em et~al.\/}(2014)Huisman, Van Der~Veen, Sun \&
  Lohse]{Huisman2014}
{\sc \au{Huisman, S.~G.}, \au{Van Der~Veen, R.~C.}, \au{Sun, C.} \& \au{Lohse,
  D.}} \yr{2014}  \at{{Multiple states in highly turbulent Taylor--Couette
  flow}}.  \jt{Nat. Commun.}  \bvol{5},  \pg{3820}.

\bibitem[Hutchins \& Marusic(2007{\natexlab{{\em a\/}}})]{Hutchins2007}
{\sc \au{Hutchins, N.} \& \au{Marusic, I.}} \yr{2007{\natexlab{{\em a\/}}}}
  \at{Evidence of very long meandering features in the logarithmic region of
  turbulent boundary layers}.  \jt{J. Fluid Mech.}  \bvol{579},  \pg{1--28}.

\bibitem[Hutchins \& Marusic(2007{\natexlab{{\em b\/}}})]{Hutchins2007large}
{\sc \au{Hutchins, N.} \& \au{Marusic, I.}} \yr{2007{\natexlab{{\em b\/}}}}
  \at{{Large-scale influences in near-wall turbulence}}.  \jt{Philos. Trans.
  Royal Soc. A}  \bvol{365}~(1852),  \pg{647--664}.

\bibitem[Kerr(2001)]{Kerr2001}
{\sc \au{Kerr, R.~M.}} \yr{2001}  \at{{Energy budget in Rayleigh-B{\'e}nard
  convection}}.  \jt{Phys. Rev. Lett.}  \bvol{87}~(24),  \pg{244502}.

\bibitem[Krug {\em et~al.\/}(2019)Krug, Baars, Hutchins \& Marusic]{Krug2019}
{\sc \au{Krug, D.}, \au{Baars, W.~J.}, \au{Hutchins, N.} \& \au{Marusic, I.}}
  \yr{2019}  \at{{Vertical coherence of turbulence in the atmospheric surface
  layer: Connecting the hypotheses of Townsend and Davenport}}.
  \jt{Bound.-Layer Meteorol.}  \bvol{172}~(2),  \pg{199--214}.

\bibitem[Lee \& Moser(2018)]{Lee2018}
{\sc \au{Lee, M.} \& \au{Moser, R.~D.}} \yr{2018}  \at{{Extreme-scale motions
  in turbulent plane Couette flows}}.  \jt{J. Fluid Mech.}  \bvol{842},
  \pg{128--145}.

\bibitem[Lohse \& Xia(2010)]{Lohse2010}
{\sc \au{Lohse, D.} \& \au{Xia, K.-Q.}} \yr{2010}  \at{{Small-scale properties
  of turbulent Rayleigh-B\'{e}nard convection}}.  \jt{Annu. Rev. Fluid Mech.}
  \bvol{42},  \pg{335--364}.

\bibitem[Marusic \& Monty(2019)]{Marusic2019}
{\sc \au{Marusic, I.} \& \au{Monty, J.~P.}} \yr{2019}  \at{Attached eddy model
  of wall turbulence}.  \jt{Annu. Rev. Fluid Mech.}  \bvol{51},  \pg{49--74}.

\bibitem[Morris {\em et~al.\/}(1993)Morris, Bodenschatz, Cannell \&
  Ahlers]{Morris1993}
{\sc \au{Morris, S.~W.}, \au{Bodenschatz, E.}, \au{Cannell, D.~S.} \&
  \au{Ahlers, G.}} \yr{1993}  \at{{Spiral defect chaos in large aspect ratio
  Rayleigh-B{\'e}nard convection}}.  \jt{Phys. Rev. Lett.}  \bvol{71}~(13),
  \pg{2026}.

\bibitem[Pandey {\em et~al.\/}(2018)Pandey, Scheel \& Schumacher]{Pandey2018}
{\sc \au{Pandey, A.}, \au{Scheel, J.~D.} \& \au{Schumacher, J.}} \yr{2018}
  \at{{Turbulent superstructures in Rayleigh-B{\'e}nard convection}}.  \jt{Nat.
  Commun.}  \bvol{9}~(1),  \pg{2118}.

\bibitem[Parodi {\em et~al.\/}(2004)Parodi, von Hardenberg, Passoni, Provenzale
  \& Spiegel]{Parodi2004}
{\sc \au{Parodi, A.}, \au{von Hardenberg, J.}, \au{Passoni, G.},
  \au{Provenzale, A.} \& \au{Spiegel, E.~A.}} \yr{2004}  \at{Clustering of
  plumes in turbulent convection}.  \jt{Phys. Rev. Lett.}  \bvol{92}~(19),
  \pg{194503}.

\bibitem[Perry \& Chong(1982)]{Perry1982}
{\sc \au{Perry, A.~E.} \& \au{Chong, M.~S.}} \yr{1982}  \at{On the mechanism of
  wall turbulence}.  \jt{J. Fluid Mech.}  \bvol{119},  \pg{173--217}.

\bibitem[van~der Poel {\em et~al.\/}(2015)van~der Poel, Ostilla-M\'onico,
  Donners \& Verzicco]{poe15c}
{\sc \au{van~der Poel, E.~P.}, \au{Ostilla-M\'onico, R.}, \au{Donners, J.} \&
  \au{Verzicco, R.}} \yr{2015}  \at{A pencil distributed finite difference code
  for strongly turbulent wall-bounded flows}.  \jt{Computers $\&$ Fluids}
  \bvol{116},  \pg{10--16}.

\bibitem[Sakievich {\em et~al.\/}(2016)Sakievich, Peet \& Adrian]{sak16}
{\sc \au{Sakievich, P.~J.}, \au{Peet, Y.~T.} \& \au{Adrian, R.~J.}} \yr{2016}
  \at{Large-scale thermal motions of turbulent {{Rayleigh-B\'enard}} convection
  in a wide aspect-ratio cylindrical domain}.  \jt{Int. J. Heat Mass Transf.}
  \bvol{61},  \pg{183--196}.

\bibitem[Shishkina {\em et~al.\/}(2010)Shishkina, Stevens, Grossmann \&
  Lohse]{shi10}
{\sc \au{Shishkina, O.}, \au{Stevens, R. J. A.~M.}, \au{Grossmann, S.} \&
  \au{Lohse, D.}} \yr{2010}  \at{Boundary layer structure in turbulent thermal
  convection and its consequences for the required numerical resolution}.
  \jt{New J. Phys.}  \bvol{12},  \pg{075022}.

\bibitem[Shishkina \& Wagner(2005)]{shi05}
{\sc \au{Shishkina, O.} \& \au{Wagner, C.}} \yr{2005}  \at{A fourth order
  accurate finite volume scheme for numerical simulations of turbulent
  {{Rayleigh-B\'enard}} convection in cylindrical containers}.  \jt{C. R.
  Mecanique}  \bvol{333},  \pg{17--28}.

\bibitem[Shishkina \& Wagner(2006)]{shi06}
{\sc \au{Shishkina, O.} \& \au{Wagner, C.}} \yr{2006}  \at{Analysis of thermal
  dissipation rates in turbulent {{Rayleigh-B\'enard}} convection}.  \jt{J.
  Fluid Mech.}  \bvol{546},  \pg{51--60}.

\bibitem[Shishkina \& Wagner(2007)]{shi07}
{\sc \au{Shishkina, O.} \& \au{Wagner, C.}} \yr{2007}  \at{Local heat fluxes in
  turbulent {{Rayleigh-B\'enard}} convection}.  \jt{Phys Fluids}  \bvol{19},
  \pg{085107}.

\bibitem[Stevens {\em et~al.\/}(2018)Stevens, Blass, Zhu, Verzicco \&
  Lohse]{stevens2018}
{\sc \au{Stevens, R. J. A.~M.}, \au{Blass, A.}, \au{Zhu, X.}, \au{Verzicco, R.}
  \& \au{Lohse, D.}} \yr{2018}  \at{{Turbulent thermal superstructures in
  Rayleigh-B{\'e}nard convection}}.  \jt{Phys. Rev. Fluids}  \bvol{3}~(4),
  \pg{041501}.

\bibitem[Stevens {\em et~al.\/}(2011)Stevens, Lohse \& Verzicco]{ste11}
{\sc \au{Stevens, R. J. A.~M.}, \au{Lohse, D.} \& \au{Verzicco, R.}} \yr{2011}
  \at{{{Prandtl}} and {{Rayleigh}} number dependence of heat transport in high
  {{Rayleigh}} number thermal convection}.  \jt{J. Fluid Mech.}  \bvol{688},
  \pg{31--43}.

\bibitem[Stevens {\em et~al.\/}(2010)Stevens, Verzicco \& Lohse]{ste10}
{\sc \au{Stevens, R. J. A.~M.}, \au{Verzicco, R.} \& \au{Lohse, D.}} \yr{2010}
  \at{Radial boundary layer structure and {{Nusselt}} number in
  {{Rayleigh-B\'enard}} convection}.  \jt{J. Fluid Mech.}  \bvol{643},
  \pg{495--507}.

\bibitem[Sun {\em et~al.\/}(2005{\natexlab{{\em a\/}}})Sun, Ren, Song \&
  Xia]{Sun2005}
{\sc \au{Sun, C.}, \au{Ren, L.-Y.}, \au{Song, H.} \& \au{Xia, K.-Q.}}
  \yr{2005{\natexlab{{\em a\/}}}}  \at{{Heat transport by turbulent
  Rayleigh--B{\'e}nard convection in 1 m diameter cylindrical cells of widely
  varying aspect ratio}}.  \jt{J. Fluid Mech.}  \bvol{542},  \pg{165--174}.

\bibitem[Sun {\em et~al.\/}(2005{\natexlab{{\em b\/}}})Sun, Ren, Song \&
  Xia]{sun05e}
{\sc \au{Sun, C.}, \au{Ren, L.-Y.}, \au{Song, H.} \& \au{Xia, K.-Q.}}
  \yr{2005{\natexlab{{\em b\/}}}}  \at{Heat transport by turbulent
  {{Rayleigh-B\'enard}} convection in 1m diameter cylindrical cells of widely
  varying aspect ratio}.  \jt{J. Fluid Mech.}  \bvol{542},  \pg{165--174}.

\bibitem[Togni {\em et~al.\/}(2015)Togni, Cimarelli \& De~Angelis]{Togni2015}
{\sc \au{Togni, R.}, \au{Cimarelli, A.} \& \au{De~Angelis, E.}} \yr{2015}
  \at{{Physical and scale-by-scale analysis of Rayleigh--B{\'e}nard
  convection}}.  \jt{J. Fluid Mech.}  \bvol{782},  \pg{380--404}.

\bibitem[Townsend(1976)]{Townsend1976}
{\sc \au{Townsend, A.~A.}} \yr{1976} {\em The Structure of Turbulent Shear
  Flow\/}.  \publ{Cambridge Univ. Press}.

\bibitem[Verma(2018)]{Verma2018}
{\sc \au{Verma, M.~K.}} \yr{2018} {\em {Physics of Buoyant Flows: From
  Instabilities to Turbulence}\/}.  \publ{World Scientific}.

\bibitem[Verzicco \& Orlandi(1996)]{ver96}
{\sc \au{Verzicco, R.} \& \au{Orlandi, P.}} \yr{1996}  \at{A finite-difference
  scheme for three-dimensional incompressible flow in cylindrical coordinates}.
   \jt{J. Comput. Phys.}  \bvol{123},  \pg{402--413}.

\bibitem[Von~Hardenberg {\em et~al.\/}(2008)Von~Hardenberg, Parodi, Passoni,
  Provenzale \& Spiegel]{Hardenberg2008}
{\sc \au{Von~Hardenberg, J.}, \au{Parodi, A.}, \au{Passoni, G.},
  \au{Provenzale, A.} \& \au{Spiegel, E.~A.}} \yr{2008}  \at{{Large-scale
  patterns in Rayleigh--B{\'e}nard convection}}.  \jt{Physics Letters A}
  \bvol{372}~(13),  \pg{2223--2229}.

\bibitem[Wang {\em et~al.\/}(2016)Wang, He \& Tong]{Wang2016}
{\sc \au{Wang, Y.}, \au{He, X.} \& \au{Tong, P.}} \yr{2016}  \at{{Boundary
  layer fluctuations and their effects on mean and variance temperature
  profiles in turbulent Rayleigh-B{\'e}nard convection}}.  \jt{Phys. Rev.
  Fluids}  \bvol{1}~(8),  \pg{082301}.

\bibitem[Young {\em et~al.\/}(2002)Young, Kristovich, Hjelmfelt \&
  Foster]{Young2002}
{\sc \au{Young, G.~S.}, \au{Kristovich, D. A.~R.}, \au{Hjelmfelt, M.~R.} \&
  \au{Foster, R.~C.}} \yr{2002}  \at{{Rolls, streets, waves, and more: A review
  of quasi-two-dimensional structures in the atmospheric boundary layer}}.
  \jt{Bull. Am. Meteorol. Soc.}  \bvol{83}~(7),  \pg{997--1002}.

\bibitem[Zhou {\em et~al.\/}(2012)Zhou, Liu, Li \& Zhong]{Zhou2012}
{\sc \au{Zhou, Q.}, \au{Liu, B.-F.}, \au{Li, C.-M.} \& \au{Zhong, B.-C.}}
  \yr{2012}  \at{{Aspect ratio dependence of heat transport by turbulent
  Rayleigh--B{\'e}nard convection in rectangular cells}}.  \jt{J. Fluid Mech.}
  \bvol{710},  \pg{260--276}.

\bibitem[Zhu {\em et~al.\/}(2018)Zhu, Phillips, Arza, Donners, Ruetsch, Romero,
  Ostilla-M\'onico, Yang, Lohse, Verzicco, Fatica \& Stevens]{zhu18b}
{\sc \au{Zhu, X.}, \au{Phillips, E.}, \au{Arza, V.~S.}, \au{Donners, J.},
  \au{Ruetsch, G.}, \au{Romero, J.}, \au{Ostilla-M\'onico, R.}, \au{Yang, Y.},
  \au{Lohse, D.}, \au{Verzicco, R.}, \au{Fatica, M.} \& \au{Stevens, R. J.
  A.~M.}} \yr{2018}  \at{{{AFiD-GPU}}: a versatile {{Navier-Stokes}} solver for
  wall-bounded turbulent flows on {{GPU}} clusters}.  \jt{Comput. Phys.
  Commun.}  \bvol{229},  \pg{199--210}.

\end{thebibliography}

\end{document}